\documentclass{article}
\usepackage{spconf,amsmath,graphicx,url}
\usepackage{hyperref}

\usepackage{fancyhdr}
\title{Prosodic clustering for phoneme-level prosody control in end-to-end speech synthesis}
\name{
\begin{tabular}{c}
	Alexandra Vioni$^{\star1}$,
	Myrsini Christidou$^{\star1}$,\thanks{$^1$Equal contribution}
	Nikolaos Ellinas$^{\star}$ ,
	Georgios Vamvoukakis$^{\star}$, \\
	Panos Kakoulidis$^{\star}$,
	Taehoon Kim$^{\dagger}$,
	June Sig Sung$^{\dagger}$,
	Hyoungmin Park$^{\dagger}$, \\
	Aimilios Chalamandaris$^{\star}$,
	Pirros Tsiakoulis$^{\star}$
\end{tabular}
}

\address{$^{\star}$ Innoetics, Samsung Electronics, Greece \\
	$^{\dagger}$ Mobile Communications Business, Samsung Electronics, Republic of Korea}

\makeindex

\begin{document}
\ninept
\maketitle
\begin{abstract}

This paper presents a method for controlling the prosody at the phoneme level in an autoregressive attention-based text-to-speech system.
Instead of learning latent prosodic features with a variational framework as is commonly done,
we directly extract phoneme-level F0 and duration features from the speech data in the training set.
Each prosodic feature is discretized using unsupervised clustering
in order to produce a sequence of prosodic labels for each utterance.
This sequence is used in parallel to the phoneme sequence in order to condition the decoder with the utilization of a prosodic encoder and a corresponding attention module.
Experimental results show that the proposed method retains the high quality of generated speech, while allowing phoneme-level control of F0 and duration.
By replacing the F0 cluster centroids with musical notes, the model can also provide control over the note and octave within the range of the speaker.
\end{abstract}

\section{Introduction}
\label{sec:intro}

Expressive speech synthesis has been of major research interest after the establishment of neural text-to-speech (TTS) systems, such as Tacotron \cite{wang2017tacotron,shen2018natural}.
Due to the high quality and naturalness of the synthesized voice, it has become possible to investigate more detailed approaches, focusing on speaker identity, speaking style, prosody control and even singing synthesis.

The task of integrating prosodic control mechanisms in neural end-to-end speech synthesis has been in the limelight, as extensive research is conducted to increase the controllability and the expressiveness of the synthesized speech.
Basic neural TTS systems implicitly model prosody and their results represent the average speaking style in the training data.
Hence, extensions of the original architectures were introduced, either to perform prosody transfer from a provided reference audio \cite{Skerry-Ryan2018} or to manually control prosody on an utterance level \cite{GSTs}.
The latter introduced the notion of the style embedding, which emerges as the weighted sum of Global Style Tokens (GSTs), a codebook which is learned in an unsupervised way.
Our work is based on these ideas and their extensions, which allow not only to control prosody on a fine-grained level, but to also utilize intuitive features to simplify the learning process.

\subsection{Related work}
\label{sec:related}

As an alternative to GSTs, Variational Autoencoders (VAEs) have also been used to learn latent representations of prosody in an unsupervised manner \cite{hsu2018hierarchical,battenberg2019effective,Zhang2019}.
While the aforementioned system variations permit prosody control only in a global sense, fine-grained prosody control has also become possible by introducing temporal structures in the prosody embedding networks, which allow pitch and amplitude control at frame-level and phoneme-level resolutions \cite{Lee2019}.
Furthermore, a hierarchical, multi-level, fine-grained VAE structure is proposed in \cite{Sun2020}, modeling word-level and phoneme-level prosody features, while a similar VAE structure with the addition of a quantization step applied to the latent vectors was adopted in \cite{Sun2020a}.

Instead of providing the Mel spectrogram of the reference audio as input to the reference encoder or variational framework, as is the case for all the systems mentioned above, specific prosodic features extracted from the reference audio, such as F0, duration and loudness, can be used as input to prosody embedding networks.
These prosodic features and their statistics can be extracted at utterance-level \cite{Shechtman2019, Gururani2019, Raitio2020} or at frame-level and phoneme-level \cite{Wan2019, Park2019} to achieve utterance-level or fine-grained prosody control, respectively.
A semi-supervised approach utilizing both Mel spectrograms and prosodic features as inputs to a variational framework is proposed in \cite{Habib2019}.
In a similar approach to ours \cite{Klimkov2019}, aggregated continuous prosodic features (F0, mgc0, duration) are used for fine-grained prosody transfer.
We differentiate our work by introducing discrete representations for arbitrary prosody control, as well as a method for disentanglement of phonetic and prosodic content.

\vspace{-5pt}
\subsection{Proposed method}

In this paper, we introduce a method for controlling prosody at the phoneme-level with discrete labels.
In similar work \cite{Sun2020a} it is shown that using a discrete prosody representation increases naturalness, while maintaining appropriate diversity.
Though, instead of utilizing a quantized fine-grained VAE, we follow a simpler approach by using intuitive features such as F0 and phoneme duration and by discretizing them with a simple clustering method.
This results in humanly interpretable labels and is directly applied to the dataset without requiring training.
We follow prior work on the end-to-end acoustic model \cite{lpctron} which is based on the Tacotron architecture \cite{wang2017tacotron,shen2018natural,gmmattention} and we extend it with additional encoder and attention modules which process the prosodic sequence.

The unsupervised K-Means algorithm is applied in order to cluster the F0 and duration information of each phoneme and capture their different levels within the speaker range.
The resulting cluster centroids form a vocabulary of discrete prosodic labels, which is used to produce a sequence of learnable prosody embeddings in parallel with the phoneme input to the acoustic model.
This proposed method enables guiding the F0 or duration of synthesized speech at a fine-grained level for the whole utterance or a specific word or phoneme by modifying their respective prosodic label, without significantly affecting naturalness.
The ability to have a discrete control sequence parallel to the phonetic input is also very intuitive because it is interpretable by the human perception and allows for straightforward manual customization.
Finally, instead of simply concatenating the prosody embeddings with the encoder outputs as is usually done, our contributions also include conditioning the decoder with an additional attention module in order to separate the phonetic and prosodic information flow during training.
The architecture with the separate phonetic encoder and attention modules allows different lengths between the prosodic and the phonetic sequences. We chose the phoneme as the unit for prosodic feature extraction,
though the proposed method can be easily adapted to work with any other linguistic units, such as syllables or words.
\vspace{-5pt}
\section{Method}
\label{sec:method}

\subsection{Forced alignment and feature extraction}
\label{sec:align}

The linguistic inputs consist of phonemes that are produced by a front-end preprocessing module from the input text.
In order to obtain accurate alignments between the utterance and its corresponding phonetic transcription, a forced-alignment system is used \cite{raptis2016expressive}.
It is an HMM monophone acoustic model trained using flat start initialization and implemented with the HTK toolkit \cite{young2002htk}, similarly to ASR forced alignment models.

After the alignments are obtained for each utterance in the training set, the duration of each phoneme is extracted.
The word boundaries and pauses are not taken into account in the F0 and duration feature extraction process, 
although they are included in the phonetic sequence to be modeled by the acoustic model.

The F0 feature for each phoneme is produced after averaging the log-F0 values for its full duration.
For F0 extraction, a standard autocorrelation method is used \cite{boersma1993accurate}, followed by interpolation and smoothing of the contour.
We found that it is better to assign the interpolated F0 value to the unvoiced phonemes, than allowing zeros which can skew the neighboring voiced values.

\subsection{Prosodic clustering}
\label{sec:clust}

After extracting the prosodic features for the entire training set, K-means with the squared distance criterion is applied, for each feature separately.
The resulting centroids can be translated as the representative values for phoneme duration or F0 and can be used as a vocabulary of tokens.

For the duration feature, clustering is performed separately per phoneme, as phoneme classes differ substantially depending on their articulation characteristics. The most prevalent duration differences may be observed between vowels and consonants. 
Additionally, the position of a phoneme inside the utterance plays an important role in its duration and thus its categorization in our experiments. 
The most prominent effect of the position of a phoneme is the phrase final lengthening, 
i.e. if a phoneme is contained in the last syllable of a phrase, it is usually pronounced with a longer duration. 
In order to accommodate for this, we perform separate clustering of the phrase final phonemes. 

At training time, for each phoneme in an utterance, its corresponding prosodic feature is assigned to the nearest cluster centroid, resulting in a sequence of prosodic labels.
Each label is represented by an embedding vector, so that a sequence similar to the phoneme input sequence is produced, which can condition the decoder.
An overview of this procedure can be seen in Figure~\ref{fig:archit}.

\subsection{Acoustic model architecture}
\label{sec:sysarch}

\begin{figure}[t]
	\begin{minipage}[b]{1.0\linewidth}
		\centering
		\centerline{\includegraphics[width=8.5cm]{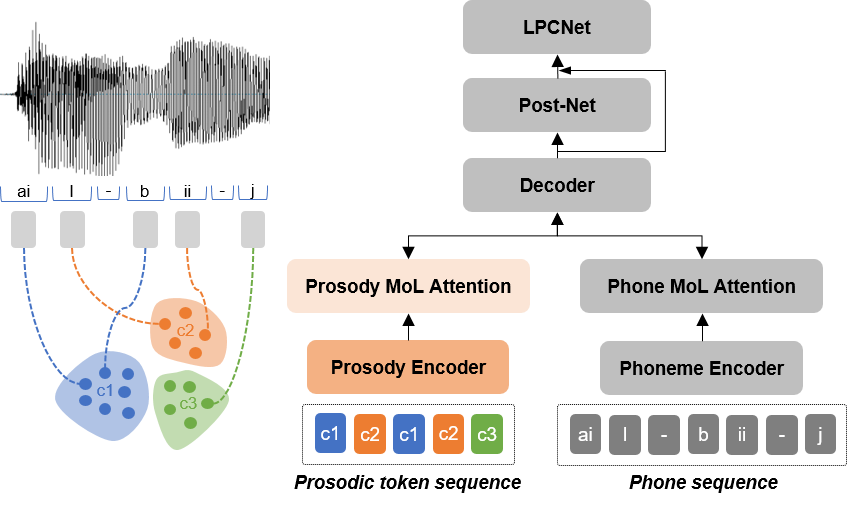}}
	\end{minipage}
	\caption{Proposed model architecture.}
	\label{fig:archit}
	\vspace{-10pt}
\end{figure}

Our work is based on the acoustic model from \cite{lpctron}. 
This model converts the input phonemes to a sequence of acoustic feature frames for the LPCNet vocoder \cite{valin2019lpcnet,srlpcnet}.
For our case, the acoustic model was enhanced with an additional encoder for processing the prosody embedding sequences.

As with the original model, the phoneme encoder converts an input sequence of phonemes $\boldsymbol{p}=[p_1,...,p_N]$ to an encoder representation $\boldsymbol{e}=[e_1,...,e_N]$. 
The prosody encoder in a similar way converts the prosody embedding sequences $\boldsymbol{p'}=[p_1',...,p_M']$ to the prosody encoder representation $\boldsymbol{e'}=[e'_1,...,e'_M]$ through a simple recurrent network.
At each decoder timestep, the attention RNN produces a hidden state $h_i$ which is used as a query in the attention mechanism for calculating the context vector $c_i$.
In our case, a secondary attention mechanism is introduced which consumes the query $h_i$ and prosody encoder representations $\boldsymbol{e'}$ and produces a prosody context vector $c_i'$.
The 2 context vectors along with the attention RNN hidden state are then fed to a stack of 2 decoder RNNs.

The new introduced prosody context vector allows the phoneme and prosody information to be modeled separately, enabling the desired fine-grained control.
A simpler approach in which the phoneme and prosody representations are directly concatenated showed worse results in terms of quality and content disentanglement.
No cluster assignment was applied to punctuation symbols or word boundary tokens, because they mainly symbolize speech pauses and we expect them to be modeled through the phoneme sequence.
As a result, the lengths of the two sequences may be different.

We expect the prosodic sequence to be parallel to the phoneme sequence in the time axis, thus requiring a robust alignment module.
For that reason we utilize the MoL attention module from \cite{lpctron} which is proven by previous research \cite{gmmattention} to maintain the monotonicity of the learned alignment, as well as to produce stable results independently of the sequence length.
This model is purely location-based and is a direct variation of the GMM attention \cite{graves2013generating}, using logistic distributions instead \cite{vasquez2019melnet}.

The Cumulative Distribution Function (\ref{cdf}) of the logistic distribution is used to compute the alignment probabilities for each decoder timestep $i$ over each encoder timestep $j$ (\ref{scores}).

\begin{equation}
	F(x;\mu,s)=\frac{1}{1+e^{-\frac{(x-\mu)}{s}}}=\sigma\left(\frac{x-\mu}{s}\right)
	\label{cdf}
\end{equation}
\begin{equation}
	a_{ij} = \sum_{k=1}^{K}w_{ik}\left(F(j+0.5;\mu_{ik},s_{ik})-F(j-0.5;\mu_{ik},s_{ik})\right)
	\label{scores}
\end{equation}

The parameters of the mixture are calculated in (\ref{muik_wik}).
\begin{equation}
	\mu_{ik}=\mu_{i-1k}+e^{\hat{\mu}_{ik}}
\quad\mathrm{}\quad
	s_{ik}=e^{\hat{s}_{ik}}
\quad\mathrm{}\quad
	w_{ik}=softmax(\hat{w}_{ik})
\label{muik_wik}
\end{equation}

The parameters $\hat{\mu}_{ik}$, $\hat{s}_{ik}$, $\hat{w}_{ik}$ are predicted by 2 fully connected layers which are applied to the attention RNN state $h_i$ as shown in (\ref{mlp}).
\begin{equation}
	\left(\hat{\mu}_{ik},\hat{s}_{ik},\hat{w}_{ik}\right)=W_2\tanh(W_1(h_i))
	\label{mlp}
\end{equation}

The context vector is calculated as the weighted sum of the encoder representations (\ref{context}).
\begin{equation}
	c_i = \sum_{j=1}^{N}a_{ij}e_j
	\label{context}
\end{equation}

The output acoustic frames are predicted by a feed-forward layer and when the decoding is complete, the prediction is finetuned by a 5-layer convolutional post-net identical to \cite{shen2018natural}. Finally, a feed-forward gate layer predicts the stop token that signals the end of speech generation.
The detailed architecture can be seen in Figure~\ref{fig:archit}.

\vspace{-5pt}
\section{Experiments and results}
\label{sec:experiments}

\begin{figure}[t]
	\begin{minipage}[b]{1.0\linewidth}
		\centering
		\centerline{\includegraphics[width=8.5cm]{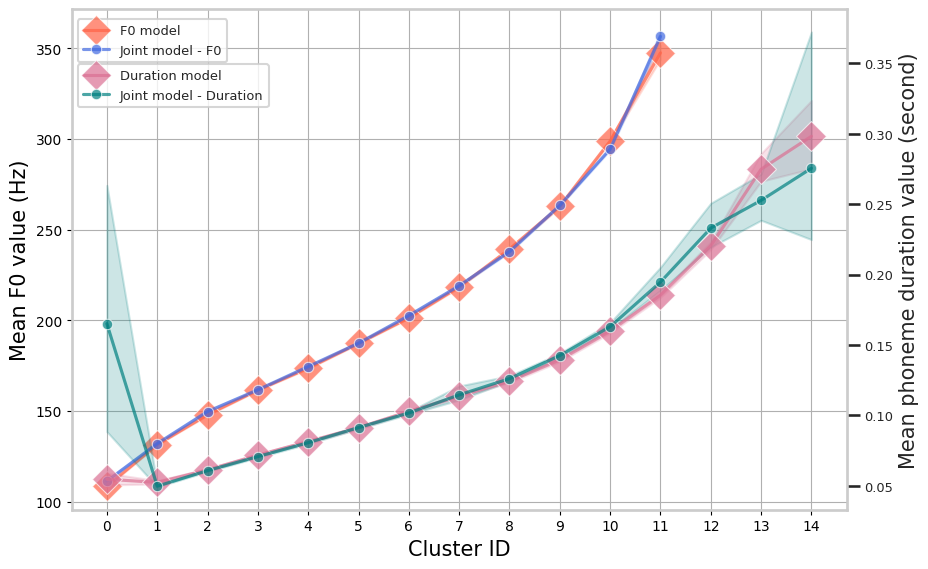}}
	\end{minipage}
	\vspace{-15pt}
	\caption{Sentence level mean F0 and average phoneme duration for ascending cluster IDs with 95\% confidence intervals.
  The left y-axis corresponds to the F0 graphs while the right y-axis corresponds to the duration graphs.}
	\label{fig:ascending}
	\vspace{-5pt}
\end{figure}

The `elbow' method \cite{joshi2013modified} was used in order to find the optimal number of clusters $k$.
The K-Means algorithm is run separately for a specified range of clusters and from the plot of a distortion metric versus $k$, the best value is selected as the inflection point of the curve.
We selected the sum of square distances to represent the distortion.
This method resulted in 12 clusters for F0 and 15 for duration.

We trained 2 separate models for F0 and duration, as well as a joint model capable of modifying both parameters.
In the joint model, the 2 sequences are represented by different prosody embedding vectors and since they have the same length, they are simply concatenated before they are passed into the prosody encoder.
The joint model is capable of modifying both parameters successfully and independently, as it is verified by the experimental results.
For the objective and subjective tests we selected 100 sentences from the dataset; those were excluded from the training and were also used to extract the ground truth prosodic labels.
The model can synthesize arbitrary text with the corresponding prosodic labels specified, predicted by a separate model, or extracted from a reference utterance.

\subsection{Objective evaluation}
\label{sec:objective}

In order to show the prosody modification capabilities of the model, we produced a test set by assigning the prosody tokens of each sentence to a single cluster in an ascending order.
For the joint model, the opposite tokens were kept at their ground truth values when not modified.
In Figure~\ref{fig:ascending} the mean values of F0 and phoneme duration are depicted, averaged over the test sentences which were modified according to a specific cluster ID.

The models are observed to follow the ascending order of the cluster IDs, verifying our hypothesis that they can modify the prosody of the speaker.
In extreme values, the performance is hindered or has high variation.
This can be accounted to fewer samples contained in these clusters which are in the extremes
of the speakers range, and are more likely to contain mislabeled data due to F0 or duration prediction errors.
In the case of the joint model, the 2 parameters can be successfully tuned separately and the acoustic results show that the modification of one parameter does not change the behavior of the second one.
We also noticed that even if a single F0 label is used for the whole utterance, the resulting prosody is adjusted but it is not flat.
This means that the phoneme embeddings also contain information about the prosody and there is not complete disentanglement, just a bias introduced by the prosody clusters which is a desired feature as it increases naturalness.
\vspace{-5pt}
\subsection{Subjective evaluation}
\label{sec:subj}

\begin{figure}[t]
	\begin{minipage}[b]{1.0\linewidth}
		\centering
		\centerline{\includegraphics[width=8.5cm]{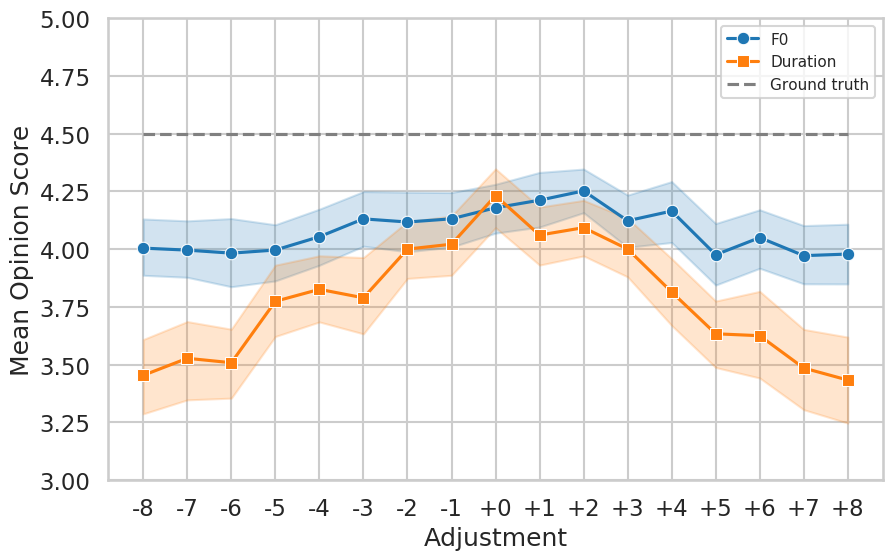}}
	\end{minipage}
	\vspace{-15pt}
	\caption{Mean opinion scores with 95\% confidence intervals.}
	\label{fig:mos}
\end{figure}

\begin{figure}[t]
	\begin{minipage}[b]{1.0\linewidth}
		\centering
		\centerline{\includegraphics[width=5cm]{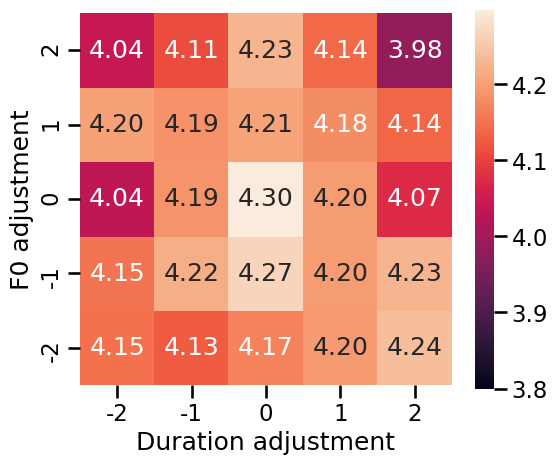}}
	\end{minipage}
	\caption{Mean opinion scores of joint model.}
	\label{fig:mosgrid}
	\vspace{-10pt}
\end{figure}

We performed listening tests in order to assess the quality of the proposed method.
The set of 100 test sentences were modified in terms of F0 and duration and the listeners were asked to score their naturalness on a 5-point Likert scale.
Considering the aim of this task, we did not introduce manual prosodic labels. 
Instead, each ground truth label is offset in the range $\left[-8,+8\right]$ for the single models and $\left[-2,+2\right]$ for the joint model in a grid manner.
We impose a limitation on the modification range because the number of samples to be scored increases significantly, especially in the joint model.
Additionally, if a label reaches the penultimate cluster ID in both positive and negative directions, then further modification for this label is halted in order to avoid the extreme centroids which were observed to sometimes be unstable. 
The resulting number of test sentences is 6000 with each sentence receiving 20 scores
by native speakers via the Amazon Mechanical Turk.

The Mean Opinion Score (MOS) is depicted as a function of the modification offset in Figure~\ref{fig:mos} for the single models and in Figure~\ref{fig:mosgrid} for the joint model.
We notice that F0 model shows less naturalness degradation and scores higher in the $+1$ and $+2$ offsets than simply feeding the ground truth prosodic labels, these small differences though are not statistically significant.
The duration model shows a clear degradation on both sides, which is attributed to the fact 
that very low or very fast speaking rate might be perceived as unnatural by some listeners.
We can also notice that the joint model scores are very high, indicating that it is capable of modifying both F0 and duration with a high output quality.
We strongly encourage the readers to listen to the samples at our website: https://innoetics.github.io
\vspace{-5pt}
\subsection{Producing musical notes}
\label{sec:notes}

A small variation of the method was also tested for producing speech that follows specific musical notes.
The corresponding musical note along with its octave are extracted from each phoneme segment according to the following formulas:

\begin{equation}
	h=\lfloor12\cdot\log_2\frac{f0}{440}\rceil+57
	\label{eq:notes1}
\end{equation}
\begin{equation}
\quad\mathrm{}\quad
	octave=\left\lfloor \frac{h}{12} \right\rfloor
\quad\mathrm{and}\quad
	note=\left(h\bmod\ 12\right)
\label{eq:notes}
\end{equation}
where \textit{h} represents the distance in semitones from the note \textit{C\textsubscript{0}}.

Instead of clustering, every distinct octave-note pair in the range of speaker is considered as a cluster centroid, and the F0 values are discretized accordingly.
On the prosody encoder side though, the octave and note information are embedded separately in order to enable modeling some pairs that may not exist in the training set.

\begin{figure}[t]
	\begin{minipage}[b]{1.0\linewidth}
		\centering
		\centerline{\includegraphics[width=\linewidth]{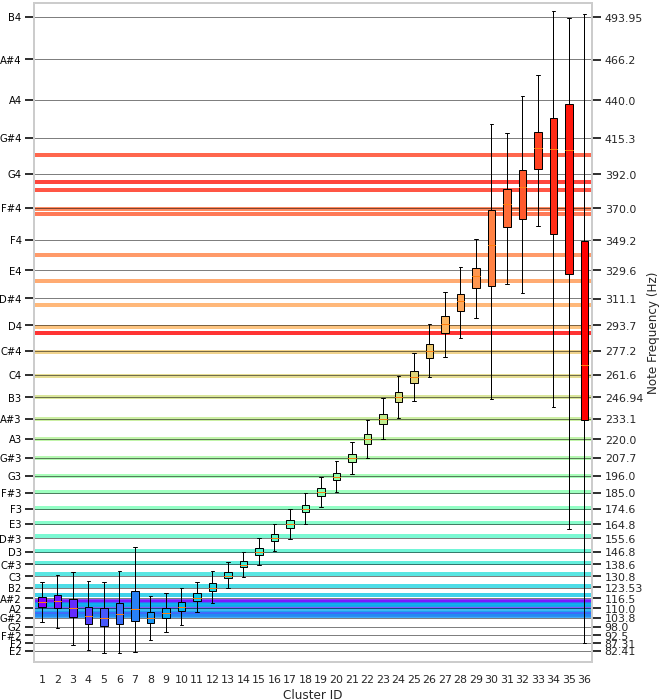}}
	\end{minipage}
	\caption{Box plot of F0 values for the musical notes production model. The colored horizontal lines show the mean F0 value for the specific cluster ID, whilst the gray lines are the note center frequencies.}
	\label{fig:notes}
	\vspace{-12pt}
\end{figure}

Results from this method are shown in Figure~\ref{fig:notes}, where the ascending progression of notes through the octaves is depicted.
Regarding the low octave, the model has similar performance for the first few notes up to G\#2, because such low F0 values are not present in the training set in a satisfying degree. Then, the F0 is modified successfully up until F\#4, after which the performance is hindered again, due to extreme F0 values, underrepresented in the training set.


\vspace{-5pt}
\subsection{Experimental setup}
\label{sec:setup}

In our experiments we use the 2013 Blizzard Challenge Catherine Byers dataset which contains 108 hours of speech. All audio data was resampled to 24 kHz.
The acoustic features were extracted in order to match the modified LPCNet vocoder \cite{srlpcnet} and consist of 20 Bark-scale cepstral coefficients, the pitch period and pitch correlation.

The phoneme encoder maps the input phoneme sequence into 256 dimensional embeddings and further applies a CBHG module.
In the prosody encoder, prosodic labels are mapped into 64 dimensional embeddings.
These are processed by a single 128-dimensional feed-forward pre-net with ReLU activation and a bidirectional GRU layer with 128 dimensions in each direction.
The decoder contains 3 recurrent layers, a 256-dimensional attention GRU and two 512-dimensional residual LSTMs.
The attention modules that are used, have a mixture of 5 logistic distributions and 256-dimensional feed-forward layers.
Dropout regularization \cite{srivastava2014dropout} of rate 0.5 is applied on all pre-net and post-net layers and zoneout \cite{zoneout} of rate 0.1 is applied on LSTM layers.

We use the Adam optimizer \cite{adam} for training the network parameters with batch size 32. The learning rate is initially $10^{-3}$ and decays linearly to $3\cdot10^{-5}$ after 100,000 iterations.
We also apply L2 regularization with factor $10^{-6}$.

\section{Conclusions}
\label{sec:conclusions}
\vspace{-2pt}
In this paper, we presented a method for creating a fully end-to-end TTS system with controllable F0 and duration at the phoneme level. 
This was achieved by preprocessing the audio data through segmentation and obtaining the duration and F0 value of each phoneme in the dataset. 
A clustering algorithm was used to separate the various F0 and durations into a number of categories, which were later used to assign learnable prosody embeddings to each phoneme. 
An additional encoder for the duration and F0 sequences as well as a separate MoL attention module were included in order to create separate alignments between the prosody encodings and the decoder hidden state, in parallel to the phoneme encoder and attention modules. 
Experimental results show that this method allows the prosody embeddings to be trained appropriately and makes fine-grained control over the F0 and duration on a phoneme level possible, without significantly affecting the naturalness of the synthetic speech.
Further work can be done in order to explore how this method can be leveraged to create multi-speaker models with wider prosodic range, increase naturalness, add emotion through duration and pitch control or even make a fully controllable singing synthesis system.
\vspace{-6pt}
\bibliographystyle{IEEEbib}
\bibliography{refs}

\end{document}